# SNR-based adaptive acquisition method for fast Fourier ptychographic microscopy

**AN PAN,**[1,2] **YAN ZHANG,**[1,2] **MAOSEN LI,**[3] **MEILING ZHOU,**[1,2] **JUNWEI MIN,**[1] **MING LEI,**[1] **AND BAOLI YAO,**[1,*]

[1]*State Key Laboratory of Transient Optics and Photonics, Xi'an Institute of Optics and Precision Mechanics, Chinese Academy of Sciences, Xi'an 710119, China*
[2]*University of Chinese Academy of Sciences, Beijing 100049, China*
[3]*Xidian University, Xi'an 710071, China*
*\*yaobl@opt.ac.cn*

**Abstract:** Fourier ptychographic microscopy (FPM) is a computational imaging technique with both high resolution and large field-of-view. However, the effective numerical aperture (NA) achievable with a typical LED panel is ambiguous and usually relies on the repeated tests of different illumination NAs. The imaging quality of each raw image usually depends on the visual assessments, which is subjective and inaccurate especially for those dark field images. Moreover, the acquisition process is really time-consuming. In this paper, we propose a SNR-based adaptive acquisition method for quantitative evaluation and adaptive collection of each raw image according to the signal-to-noise ration (SNR) value, to improve the FPM's acquisition efficiency and automatically obtain the maximum achievable NA, reducing the time of collection, storage and subsequent calculation. The widely used EPRY-FPM algorithm is applied without adding any algorithm complexity and computational burden. The performance has been demonstrated in both USAF targets and biological samples with different imaging sensors respectively, which have either Poisson or Gaussian noises model. Further combined with the sparse LEDs strategy, the number of collection images can be shorten to around 25 frames while the former needs 361 images, the reduction ratio can reach over 90%. This method will make FPM more practical and automatic, and can also be used in different configurations of FPM.

## 1. Introduction

Fourier ptychographic microscopy (FPM) [1-5] is a recently proposed computational imaging technique which has found wide applications in the field of digital pathology. FPM approach overcomes an inherent trade-off between high resolution (HR) and large field-of-view (FOV) in conventional microscope, by synthesizing a series of variably-illuminated low-resolution image measurements, which is a particularly time-consuming process. In order to reduce the acquisition time, by making full use of the data redundancy, Dong et al. [6] proposed the sparse sampling method to solve the over/underexposed issues and relax the requirement of imaging sensor's dynamic range. Bian et al. [7] proposed the AFP algorithm to reduce the redundancy of the spectrum, which can be analogous to lossy compression [8]. Zhang et al. [9] proposed a self-learning method to further reduce the data collection time of AFP, which can reduce the data size over 70% without sacrificing image quality. Guo et al. [10] redesigned the LED panel to break the periodicity of traditional prototype and reduce the number of image acquisitions over 50%. Tian et al. [11] proposed a multiplexing scheme with 4 to 8 LEDs to illuminate simultaneously. Further, they successfully achieved high-speed in vitro imaging with only 21 raw images by combining the DPC [12] algorithm, obtaining a comparable reconstruction with that by 173 images [13]. All these methods improve the efficiency of FPM, but they correspondingly increase the algorithm complexity and computational burden. On the contrary, Sun et al. [14] proposed an alternative imaging strategy, which omits multiplexing and utilizes sparse LED arrays to reduce the data redundancy, may achieve better results with the same amount of images as it may suffer less from background noise and signal fluctuation. It may be deduced that sequential collecting 21 raw images with sparse LED arrays can also implement the results of ref. [13], though it hasn't been proven.

Theoretically the synthetic numerical aperture (NA) of the FPM approach can be very large, but there are some limitations in practice. On the one hand, LED intensities are declined with the increasing incident angle θ (proportional to $\cos^4\theta$) in the conventional LED-based FPM configuration [15], restricting the achievable synthetic NA within the range of 0.3–0.8. On the other hand, the dark-field (DF) images with high-angle illuminations are easily submerged by the noise due to the low signal-to-noise ratio (SNR). Some low-SNR images may import more noise and affect the final results. Proper abandonment of these low SNR images may be beneficial to the reconstruction quality. But the quantification of SNR in FPM system is rarely reported in previous publications and usually depends on the visual assessments, which is subjective and inaccurate especially for those DF images. Some DF images may contain valuable information which are not sensitive to visual perception but can be recognized by quantitative evaluation. Besides, how much effective NA can a system get is ambiguous and usually relies on the repeated tests of different illumination NAs. To this end, we propose a SNR-based adaptive acquisition method, for quantitative evaluation and adaptive collection of each raw image according to the SNR value, to improve the FPM's acquisition efficiency and also automatically obtain the maximum achievable NA under a certain experimental configuration. In order to save valuable information as much as possible and make the approach nonparametric, threshold should take the maximum SNR of the edge images, which can be acquired by some simple tests. Compared to AFP, SNR-based adaptive acquisition method is more like a kind of lossless compression technique [8].

Noise estimation is an essential step for determining the SNR value. Considering of practicability, this estimation shouldn't be too complicated. Generally, the dark current noise is one of the dominant factors due to the long exposure time of the illumination of LED panels, while the Gaussian noise and Poisson are another two dominant factors due to electrothermal motion and photon conversion efficiency [16, 17]. Fan et al. [18] proposed an adaptive denosing method for FPM, which can evaluate the noise distribution of each raw image iteratively. But it still needs a time-consuming acquisition with all illumination angles and a series of subsequent calculations. Whereas some preprocessing procedures [19, 20] are very direct and effective, though some parts of the meaningful signals are inevitably wiped out along with the noise. Note that the discussion of SNR is meaningless as an absolute number and is only useful for comparison purposes. So in this work, we propose two methods for noise evaluation. One is on the basis of maximum likelihood estimation (MLE) method to evaluate the Poisson or Gaussian noise, which is widely favored for the estimation of cost function [21, 22]. Another is based on the preprocessing procedure [19], where the noise distribution is reflected by the average of local background. Because the unknown modality of the stray lights may break the Poisson or Gaussian model and affect the estimation of MLE method. Then the higher value between the two methods is selected as the noise estimation. The algorithm in our procedure is still the widely used EPRY-FPM algorithm [23], with an excellent convergence property and a low computational complexity. Some other system correction and noise suppression algorithms can also be used to further improve the reconstruction quality [24-27]. This paper compares the imaging quality of USAF targets and biological samples with both 8-bit CCD and 16-bit CMOS respectively. In our experiment, the dark current and Gaussian noise are dominant in 8-bit CCD, while the dark current and Poisson noise are dominant in 16-bit CMOS. Further, we reduce the data redundancy with sparse LED arrays as described above to further shorten the acquisition time. Compared to 361 (19×19) images used in original scheme, 25 images for USAF targets and 22 images for rabbit tongue section are needed in our proposed method with 16-bit camera. As for 8-bit camera, the number of raw captures decreased to 21 and 28 respectively, while originally need 225 (15×15) images, significantly shorten the acquisition time (reduced by over 90%) compared with original 225 (15×15) images, without sacrificing the reconstruction quality. The approach will make FPM more practical and automatic, which can also be used in different configuration of FPM [28-30].

## 2. Method

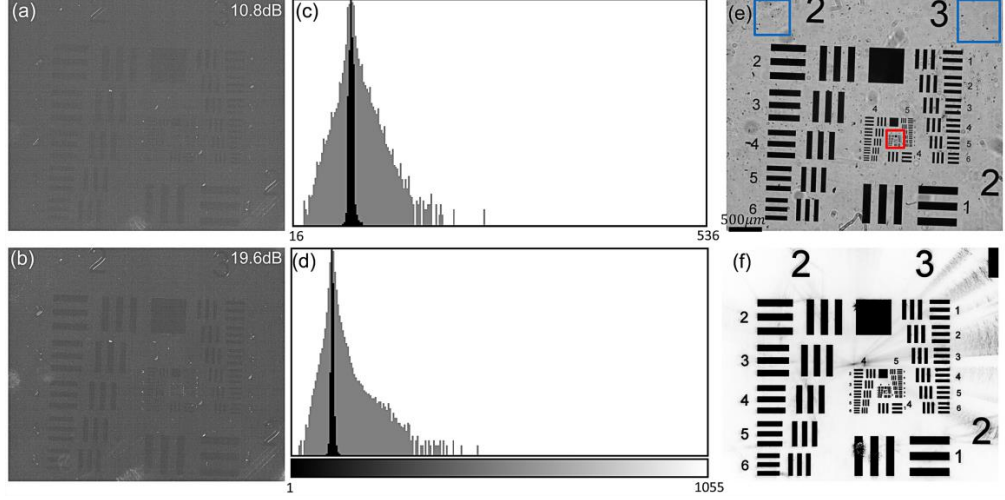

Fig.1. Illustration for the evaluation of SNR. (a) and (b) present two typical DF image at the edge of LED arrays with contrast stretching, while (c) and (d) present their histograms respectively. (e) presents one bright field image captured by a 4×/0.1NA objective, while the red rectangle is the ROI and the blue rectangles are the local background. (f) is the threshold processing of (e), which can see a little stray light.

For illustration, Fig.1(a) and 1(b) present two typical DF images at the edge of LED arrays with contrast stretching captured by a 4×/0.1NA objective and a 16-bits camera, while (c) and (d) present their histograms respectively. The gray parts in histogram are dealt with logarithmic operation. It's really hard for human eyes to decide the image quality even though with contrast stretching, for instance, whether it is contaminated by noise or stray light and whether it can be used as one of the datasets. More importantly, if the results are not very well, it's hard to decide what raw images should be in charge. As a consequence, it's necessary to introduce a quantitative criterion for image evaluation and adaptive data acquisition. SNR is frequently defined as the ratio of the signal power and the noise power. But in imaging, an alternative definition with PSNR can be found, where PSNR is given as the ratio of the maximum signal value and noise standard deviation, which can also be denoted by mean squared error (MSE) [8]. But usually in experiment, the noise image is unknown. So we define a formula to evaluate the PSNR as follows.

$$PSNR_N = 20 log_{10} \frac{I_{max} - I_n}{I_n} \qquad (1)$$

where $I_{max}$ is the maximum intensity of the region of interest (ROI) illustrated in the red rectangle of Fig.1 (e). $I_n$ is the standard deviation of the noise. The subscript $N$ of $PSNR$ is to denote the ordinal number of images. Before calculation, each raw image needs to compensate for the uneven of the intensity and a cosine function is divided for data preprocessing [15]. The PSNR of Fig.1 (a) is 10.8dB, which won't be collected for the datasets according to the threshold value 19dB, while the PSNR of Fig.1 (b) is 19.6dB, which will be used in the experiment.

Two methods are used to evaluate the noise. One is to use the MLE method to evaluate the Poisson or Gaussian noise, which has been deduced for cost function in the literature [17, 21] and will not be detailed here. In our experiment, the dark current and Gaussian noise are dominant in 8-bit CCD, while the dark current and Poisson noise are dominant in 16-bit CMOS. The noise $I_n$ can be deemed as two parts $I_n = \overline{\sigma_P} + I_D$ or $I_n = \overline{\sigma_G} + I_D$ respectively according to different noise models. The $\overline{\sigma_P}$ is the estimation of standard deviation of

Poisson noise. And $\overline{\sigma_G}$ is the estimation of standard deviation of Gaussian noise. The dark current noise $I_D$ can be estimated by averaging multiple background images without object and illuminations. In our experiment, the gray value of the dark current is generally 101 with16-bits camera and 0.2 with 8-bits camera. After subtracting the dark current noise, the residue noise can be deemed as Poisson or Gaussian distribution. And the Poisson maximum-likelihood function is as follows [17, 21]:

$$\min L_{Poisson} = \min \sum_N \sum_{x,y} \left( -I_N(x,y) \log\left(\overline{I_N}(x,y)\right) + \overline{I_N}(x,y) + \log\left(I_N(x,y)!\right) \right)$$
$$\approx \min \sum_N \sum_{x,y} \frac{\left(I_N(x,y) - \overline{I_N}(x,y)\right)^2}{2\sigma_P^2} \tag{2}$$

where $\overline{I_N}(x,y)$ is the maximum likelihood estimation of each raw image. And the standard deviation of Poisson noise $\overline{\sigma_P}$ can be approximate to:

$$\sigma_P(x,y) = \sqrt{I(x,y) - I_P(x,y) - I_D} \approx \sqrt{I(x,y) - I_D} \tag{3}$$

Then for simplicity, the $\overline{\sigma_P}$ can be the half maximum value of $\sigma_P(x,y)$. For white Gaussian noise, it's assumed that $\overline{\sigma_G}$ is the same across all pixels for all images, namely a constant, though this will not always be the case for FPM datasets. By making a Gaussian noise assumption, the Gaussian maximum-likelihood function is as follows [17, 21]:

$$\min L_{Gaussian} = \min \sum_N \sum_{x,y} \left[ \frac{1}{2} \log\left(2\pi\sigma_G^2\right) + \frac{\left((I_N(x,y)) - (\overline{I_N}(x,y))\right)^2}{2\sigma_G^2} \right]$$
$$\Leftrightarrow \min \sum_N \sum_{x,y} \left[ \left((I_N(x,y)) - (\overline{I_N}(x,y))\right)^2 \right] \tag{4}$$

Then the standard deviation of Gaussian noise $\overline{\sigma_G}$ is not associated with the object due to its nature of additive noise and can be evaluated as follows:

$$\overline{\sigma_G} = std\left(I^{'} - I_D\right) = std\left(I^{'}\right) \tag{5}$$

where $std(\cdot)$ is the operation of standard deviation and $I^{'}$ is the average of multiple bright field image without loading objects. Another method for noise estimation is to utilize the preprocessing procedures [19], where the noise distribution can be reflected by the average of local background as indicated in blue boxes in Fig.1 (e). Since the unknown modality of the stray lights may break the noise model and affect the estimation of MLE method, the local noise can be intensified as shown in Fig.1 (f). The noise level calculated by these two methods is under the same scale, which is convenient to take the relatively higher value as the noise estimation.

### 3. Experimental results

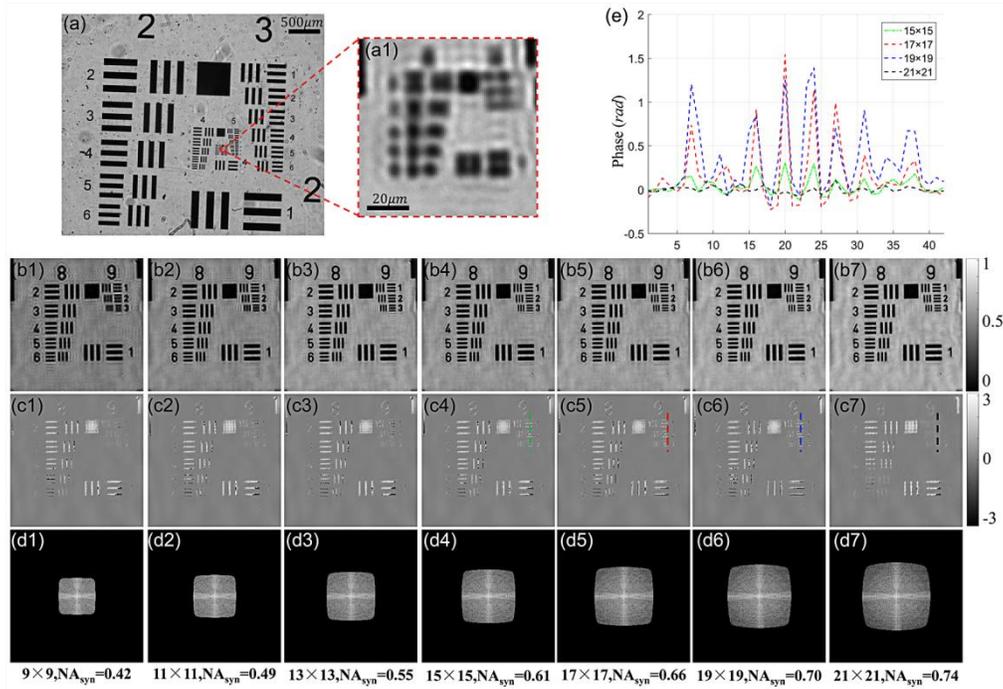

Fig.2. The recovery results of the USAF target under different illumination NA respectively with a 16-bits camera. (a) presents the FOV of the USAF target captured by a 4×/0.1NA objective. (a1) shows the enlargement of one segment (50×50 pixels) as the LR intensity image. Group (b), (c) and (d) denote the reconstructed intensity, phase and spectrum, respectively. (e) presents the profiles of group 9, element 1-3 in (c4)-(c7).

The experimental setup and data acquisition process for FPM can be found in the literature [1] and will not be detailed here. Fig.2 (a) presents the FOV of a USAF target captured by an 4×/0.1NA apochromatic objective and a 16-bits sCMOS (Neo 5.5, Andor) with the pixel size of 6.5$\mu m$. Figure 2 (a1) shows the enlargement of one segment (50×50 pixels) of Fig.2 (a), which becomes blurry since it is restricted by the low NA of objective. A 32×32 programmable LED matrix with an illumination wavelength of 631.13nm, 20nm bandwidth and 4mm spacing is placed at 67.5mm above the sample stage. Groups (b), (c) and (d) denote the reconstructed intensity, phase and spectrum, respectively. The theoretical synthetic NAs are presented below each column. The noise sources are mainly from Poisson noise and dark current noise and some preprocess methods have been used in this experiment. From Fig.2 (b1)-(b4), resolution is gradually improved with the increasing of synthetic NAs, but from Fig.2 (b4)-(b7), it's hardly to tell which one is better or worse. However, it can be seen that the phase quality of Fig.2 (c4)-(c7) is gradually improved and then deteriorated, whose profiles of group 9, element 1-3 are shown in Fig.2 (e). The blue dashed line has a better contrast than others. Thus the maximum effective NA in practice is around 0.7 with 19×19 LED arrays, which is the reason why most of the articles get the maximum NA around 0.7 [11, 13]. So it isn't the more images we acquire the better the quality we get. On the contrary, phase will be deteriorated when the synthetic NAs beyond a certain degree. Because more noise will be simultaneously introduced with the increasing of synthetic NAs. Some signal will be inevitably merged into the noise at the high angle. It may conclude that the noise will have more impact on the reconstruction of phase rather than the intensity, while the phase is more important in a phase imaging technique.

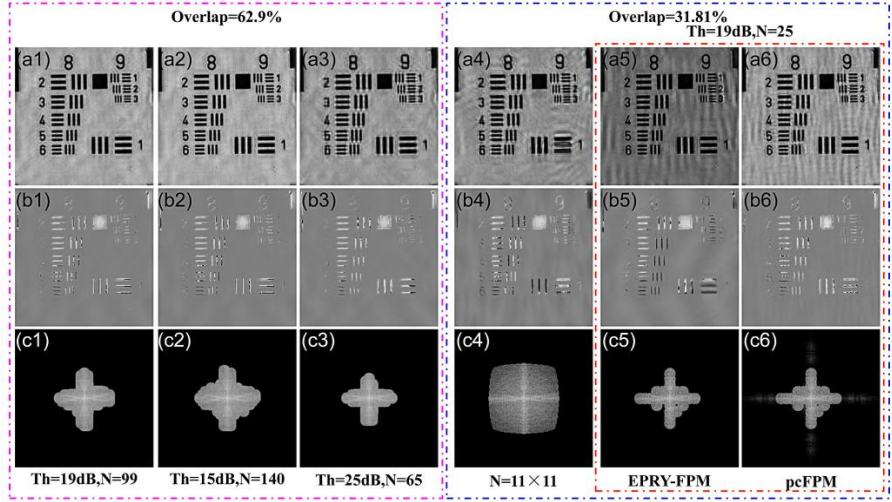

Fig.3. The recovery results of the USAF target under different threshold and overlap rate. Group (a), (b) and (c) denote the reconstructed intensity, phase and spectrum, respectively.

Considering that the shape of USAF is in the vertical and horizontal directions, the main spectral information is also in the same directions. And the SNR will ideally have the same distribution. Thus the information of four corners is not so important indeed and may be ignored due to low SNR. It's possible to ignore these images to improve the efficiency of FPM. The SNR distribution of Fig.2 (d6) is from -3dB to 47dB. If you want to save as much information as possible, the threshold should be set lower, for example the results of 15dB are shown in Fig.3 (a2), (b2) and (c2), and the synthetic NA is unchanged with 140 images, so it can be seen as lossless compression compared with Fig.2. (b6) and (c6). Some asymmetric of spectrum is the spectrum of the number in the USAF target. If the imaging speed is required, the threshold could be chosen at a higher one, for instances 25dB shown in Fig.3. (a3), (b3) and (c3) with 65 images, which is similar to the AFP that can be seen as the lossy compression. The group 9, element 2 and 3 become a little blurry. However, in order to let the process nonparametric and automatic, the threshold should be given by the maximum SNR of quadrangular edge to ensure that the synthesis NA will not change too much as shown in Fig.3 (a1), (b1) and (c1) with 99 images, which can be also approximate to lossless compression. It can be seen that the results with our methods are similar to the results of Fig.2 (b6) and (c6) but with less images. On the one hand, we are able to automatically get the maximum effective NA no longer relying on the subjective judgment and repeating tests. On the other hand, the maximum effective NAs can be unchanged, and the images required can be reduced around 73% comparing to the original 361 images. Although it may not reduce too much images, the imaging speed can be improved by other methods. Here we combine the scheme of sparse LED arrays [14] to reduce the LED redundancy by adjusting the 4mm spacing to 8mm. It's quite easy to achieve without any changes of the original setup compared with ref.[10]. And the overlap rate can be reached to the limitation of 31.81% [14]. Then we achieve quite the same results by using just 25 images, while it originally needs 121 images to synthesis 0.7 NA. Note that when the images are reduced, the stability will decline, which may be easy to suffer from the system errors as shown in Fig.3 (a5) and (b5). Compared with Fig.3 (a4) and (b4), the contrast is decreased. It's better to use the pcFPM algorithm [24] to adjust the system parameters as shown in Fig.3 (a6), (b6) and (c6). But in our experiments, we only update the shift factors $\Delta x$, $\Delta y$ along the x- and y-axis, the rotation factors $\Delta \theta$ but we don't update the height factor $h$ for robust. Because it's found that the height factor will be extremely wrong, which may not helpful the results. It's worth mention that different from AFP, the SNR not only reflect the degree of importance, but also the imaging quality in reality. Thus in some case a valuable image may not be captured if it suffer from stray light or noise too much due to low SNR. For example, in Fig.3 (c5) and (c6) some center images have be ignored. Consequently, we reach the ultimate limitation of 25 images to retrieval the USAF targets while it needs 361 images in the beginning.

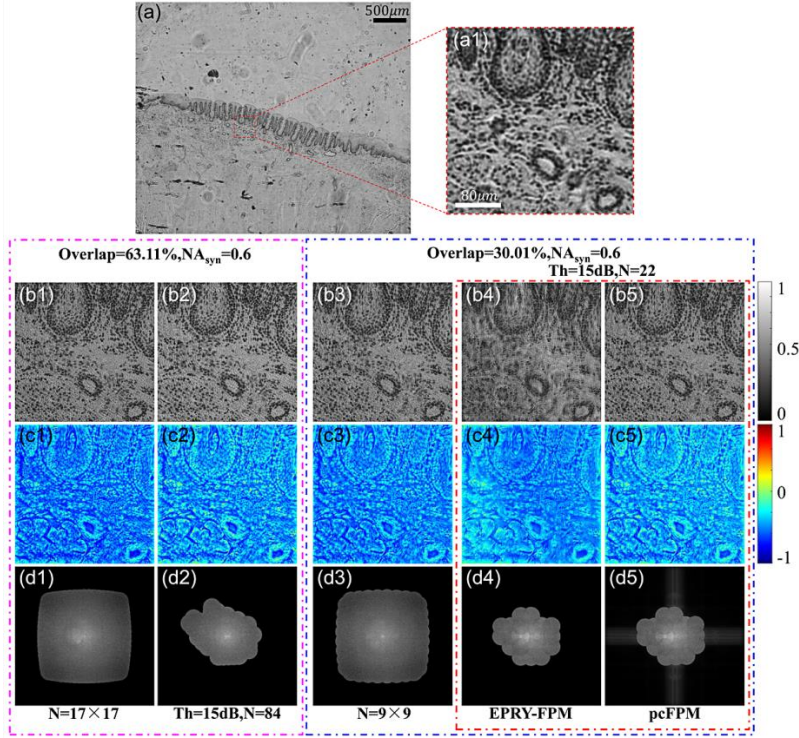

Fig.4. The recovery results of the rabbit tongue section with original FPM and SNR-based fast FPM under different overlap rates. (a) presents the FOV of the sample captured by a 4×/0.1NA objective. (a1) shows the enlargement of one segment (200×200 pixels) as the LR intensity image. Group (b), (c) and (d) denote the reconstructed intensity, phase and spectrum, respectively.

In addition, we also test our method in a biological sample (rabbit tongue section) compared with original FPM as shown in Fig.4. The LED array is placed at 67.9mm above the sample. Other parameters are the same as those above. Figure 4 (a) presents the FOV of rabbit tongue section with an 4×/0.1NA objective, while Fig.4 (a1) shows the enlargement of one segment (200×200 pixels) as the LR intensity image. Group (b), (c) and (d) denote the reconstructed intensity, phase and spectrum, respectively. Figure 4 (b1), (c1) and (d1) show the results of maximum effective NA after repeating tests, resulting in a final synthetic NA of 0.6. Compared with original 289 images shown in Fig.4 (b1), (c1) and (d1), our method needs fewer images at 84 with 15dB threshold. Note that the threshold is automatically get and is different to the threshold of USAF targets due to different absorptive capacity. Further, combined with the sparse LED arrays, we ultimately achieve 22 images to retrieval the samples. Compared with original 289 images, the reduction ratio can reach over 90%. And compared with 81 images in the sparse strategy shown in Fig. 4 (b3), (c3) and (d3), the reduction ratio can also reach over 69%.

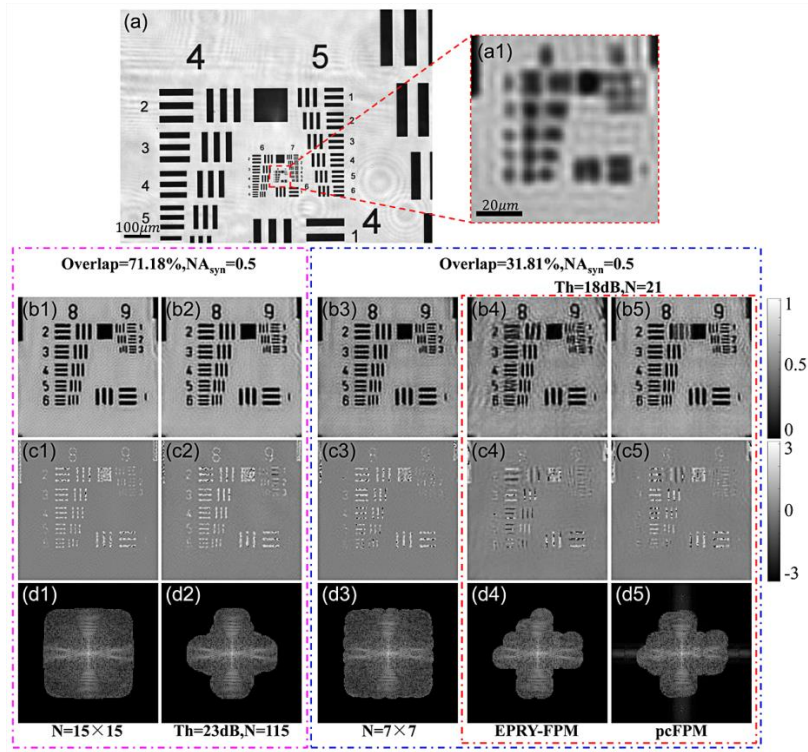

Fig.5. The recovery results of the USAF target with original FPM and SNR-based fast FPM with an 8-bits camera. (a) presents the FOV of the taget captured by a 4×/0.1NA objective. (a1) shows the enlargement of one segment (90×90 pixels) as the LR intensity image. Group (b), (c) and (d) denote the reconstructed intensity, phase and spectrum, respectively.

This method can be also used in the 8-bits cameras, while 8-bits cameras have different noise model. The main noise sources are Gaussian noise and dark current noise with our 8-bits camera. Compared with 16-bits cameras, 8-bits are much cheaper, which is good for FPM to commercialize. Fig.5 (a) presents the FOV of a USAF target captured by an 4×/0.1NA apochromatic objective and a 8-bits camera (DMK23G445, Imaging Source Inc., Germany) with the pixel size of 3.75$\mu m$. Figure 5 (a1) shows the enlargement of one segment (90×90 pixels) of Fig.5 (a), which becomes blurry since it is restricted by the low NA of objective. The LED array is placed at 87.5mm above the sample stage. Groups (b), (c) and (d) denote the reconstructed intensity, phase and spectrum, respectively. Figure 5 (b1), (c1) and (d1) present the results of maximum effective NA after repeating tests, resulting in a final synthetic NA of 0.5. It's quite smaller compared with the 16-bits cameras in Fig.2 due to the capability. The group 9, element 1-3 in Fig. 5 (b1) and (c1) are little blurry than Fig. 2 (b6) and (c6). But the contrast of phase is higher than with 16-bits cameras, which may be the contribution of high SNR indicated by the SNR threshold. Compared with 225 images in original FPM, our method only needs 115 images as shown in Fig.5 (b2), (c2) and (d2). And combined with the sparse LED arrays, we achieve 21 images to retrieve the USAF targets. In order to let the overlap rate close to the limitation, the distance between LEDs and samples is changed to 70mm. And it's worth mentioning that the SNR threshold will change when the conditions are changed.

In addition, we also test the rabbit tongue section with 8-bits cameras compared with original FPM as shown in Fig.6. The LED array is placed at 70mm above the sample. Other parameters are the same as above. Figure 6 (a) presents the FOV of rabbit tongue section with an 4×/0.1NA objective, while Fig.6 (a1) shows the enlargement of one segment (200×200 pixels) as the LR intensity image. Group (b), (c) and (d) denote the reconstructed intensity, phase and spectrum, respectively. Figure 6 (b1), (c1) and (d1) show the results of maximum effective NA after repeating tests, resulting in a final synthetic NA of 0.6. Compared with original 225 images shown in Fig.6 (b1), (c1) and (d1), our method needs fewer images at 98 at 17dB threshold. Further, combined the sparse LED arrays, we ultimately achieve 28 images to retrieval the samples. The conclusions are also the same as USAF in Fig.5 that the contrast of phase with 8-bits is better than with 16-bits cameras.

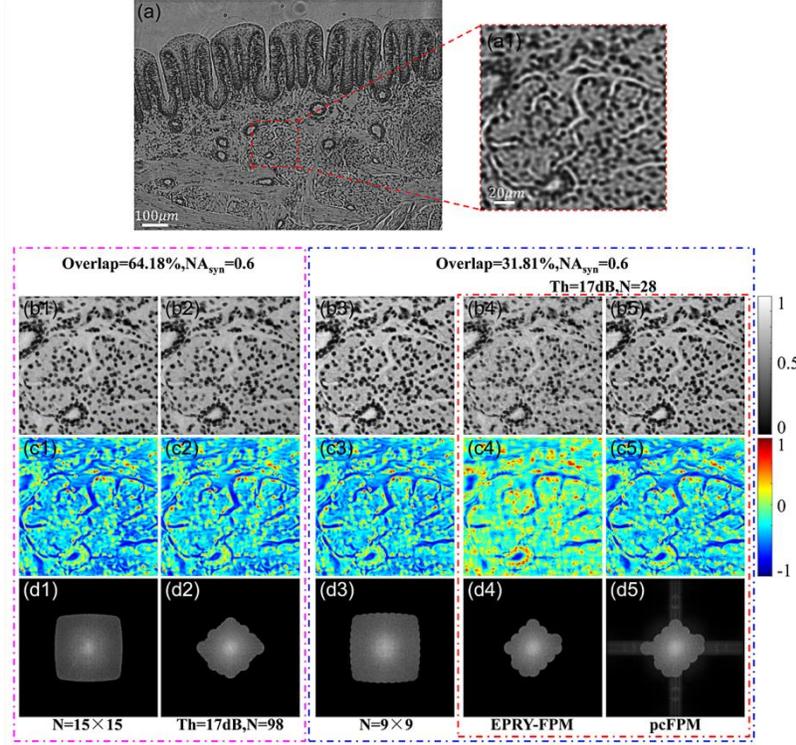

Fig.6. The recovery results of the rabbit tongue section with original FPM and SNR-based fast FPM with an 8-bits camera. (a) presents the FOV of the sample captured by a 4×/0.1NA objective. (a1) shows the enlargement of one segment (200×200 pixels) as the LR intensity image. Group (b), (c) and (d) denote the reconstructed intensity, phase and spectrum, respectively.

## 4. Conclusion and discussion

We have proposed a SNR-based adaptive acquisition method for fast Fourier ptychographic microscopy, which can evaluate each raw image objectively without the assessment of subjective vision and can automatically ignore those low SNR images to reduce the time of collection, storage and subsequent calculation. In order to improve the efficiency of FPM further, we combine the sparse LED scheme without any changes of the original LED arrays to achieve 25 images and 21 images, while it needs 361 images in 16-bits cameras and 225 images in 8-bits cameras, respectively. The reduction ratio can reach over 90%. This method is simple and effective to both USAF targets and biological samples, which can also be used in different configurations of FPM, such as the reflection, fluorescence and so on.

Besides, two open questions are worth to investigating. First, the intensity of LEDs in our experiments is still too weak. The exposure time with 16-bits cameras is 0.7s per frame. Thus the sum exposure time is around 252.7s for 361 images, although only 25 images are needed. If the intensity improves, for instance, 40ms per frame in some literature [5], then the time can be 14.44s. And by changing the sequence of lighting LED from center to edge and automatically skipping those low SNR images according to the trend, the collection time can be reduced further. Second, from the experimental results, it theoretically exists an around 25 LED array to automatically achieve fast FPM. But it will be difficult to design a specific 25 LED array because the optimal LED locations of different samples are not always the same. If the results of using such a unique LED array satisfy the requirements, it will further reduce the exposure time to less than 1s. Then the dynamic imaging can be achieved and may quicker than the fastest record of 1.25Hz.

**Funding**